\documentclass[letterpaper]{article} 
\usepackage{aaai2026}  
\usepackage{times}  
\usepackage{helvet}  
\usepackage{courier}  
\usepackage[hyphens]{url}  
\usepackage{graphicx} 
\usepackage{natbib}  
\usepackage{caption} 
\usepackage{algorithm}
\usepackage{algorithmic}

\usepackage{newfloat}
\usepackage{listings}

\usepackage{amsfonts}
\usepackage{bm}
\usepackage{amsmath} 
\usepackage{multirow}
\usepackage{booktabs}
\usepackage[table]{xcolor}
\usepackage{makecell} 
\usepackage{array}    
\DeclareCaptionStyle{ruled}{labelfont=normalfont,labelsep=colon,strut=off} 
\floatstyle{ruled}
\newfloat{listing}{tb}{lst}{}
\floatname{listing}{Listing}
\title{BLADE: A Behavior-Level Data Augmentation Framework with Dual Fusion Modeling for Multi-Behavior Sequential Recommendation}

\author{
Yupeng Li\textsuperscript{\rm 1},
Mingyue Cheng\textsuperscript{\rm 1}\thanks{Corresponding Author.},
Yucong Luo\textsuperscript{\rm 1},
Yitong Zhou\textsuperscript{\rm 1},
Qingyang Mao\textsuperscript{\rm 1},
Shijin Wang\textsuperscript{\rm 2} 
}

\affiliations{
\textsuperscript{\rm 1}State Key Laboratory of Cognitive Intelligence, University of Science and Technology of China\\
\textsuperscript{\rm 2}iFLYTEK AI Research (Central China), iFLYTEK Co., Ltd.\\
\{liyupeng,prime666,yitong.zhou,maoqy0503\}@mail.ustc.edu.cn, mycheng@ustc.edu.cn, sjwang3@iflytek.com
}

\usepackage{bibentry}

\begin{document}

\maketitle

\begin{abstract}
Multi-behavior sequential recommendation aims to capture users' dynamic interests by modeling diverse types of user interactions over time. Although several studies have explored this setting, the recommendation performance remains suboptimal, mainly due to two fundamental challenges: the heterogeneity of user behaviors and data sparsity. To address these challenges, we propose BLADE, a framework that enhances multi-behavior modeling while mitigating data sparsity. Specifically, to handle behavior heterogeneity, we introduce a dual item-behavior fusion architecture that incorporates behavior information at both the input and intermediate levels, enabling preference modeling from multiple perspectives. To mitigate data sparsity, we design three behavior-level data augmentation methods that operate directly on behavior sequences rather than core item sequences. These methods generate diverse augmented views while preserving the semantic consistency of item sequences. These augmented views further enhance representation learning and generalization via contrastive learning. Experiments on three real-world datasets demonstrate the effectiveness of our approach.
\end{abstract}
\begin{links}
    \link{Code}{https://github.com/WindSighiii/BLADE}
\end{links}

\section{Introduction}
Recommender systems have become essential components of online platforms such as e-commerce and social media, enabling personalized content delivery and alleviating information overload \cite{he2017neural,exp,he2023survey,cheng2025}. Among various recommendation paradigms, sequential recommendation (SR) has received substantial attention due to its capability to model the temporal dependencies within user-item interactions and predict subsequent user actions \cite{fang2020deep,wang2019sequential,fpmc,cheng2021,cheng2022,timer1,timereasoner}. To better capture complex user intentions, recent studies have extended SR to multi-behavior sequential recommendation (MBSR), which incorporates diverse user behaviors (e.g., clicks, favorites, and cart additions) to assist purchase predictions in e-commerce contexts \cite{nextip,chain,denoise}. Further extending this concept, EIDP \cite{chen2024explicit} introduced Behavior Set-informed Sequential Recommendation (BSSR), where each interaction is represented as a set of concurrent behaviors (e.g., likes, favorites, and shares) on the same item. This formulation enables more expressive and fine-grained modeling of user preferences, making it particularly suitable for scenarios like social media where the rich co-occurrence of behaviors is common.

Despite its enhanced modeling capacity, BSSR still faces two key challenges. First, different behavior types have distinct semantics and complex dependencies, and the behavior set format further increases the difficulty, making behavior heterogeneity a greater challenge in modeling user preferences. To handle such heterogeneity, existing methods have proposed several item-behavior fusion strategies: (1) Early fusion integrates item and behavior representations at the input level, enriching local semantics but potentially introducing semantic interference due to differences in representation spaces \cite{rib,BINN,he2022bar}; (2) Intermediate fusion introduces behavioral information in intermediate layers to mitigate interference, but the final user representation remains item-dominated, missing the collaborative effect brought by direct item-behavior fusion \cite{mbstr}; (3) Late fusion models each behavior separately as sub-sequences before aggregation, preserving independence but failing to effectively model cross-interactions among items and behaviors \cite{dumys}. Second, the inherent data sparsity limits the model's ability to learn reliable representations. Although self-supervised learning and data augmentation have proven effective in mitigating data sparsity in SR \cite{cl4rec, duorec, ICLRec} and MBSR \cite{xiao2024generic}, these methods are primarily designed for item sequences. However, item sequences encode the core semantics of user preferences. Operating on item sequences directly may distort the core semantics, causing the contrastive views to deviate from users’ true preferences and ultimately weakening the learned representations. 

To address these challenges, we propose BLADE, a \textbf{B}ehavior-\textbf{L}evel data \textbf{A}ugmentation framework with \textbf{D}ual fusion
mod\textbf{E}ling, which enhances multi-behavior modeling and simultaneously mitigates data sparsity. Specifically, we first introduce a dual item-behavior fusion architecture to handle behavior heterogeneity by combining early and intermediate fusion, capturing both static and dynamic item-behavior relations to enhance multi-behavior semantic modeling. Then, we design three behavior-level data augmentation methods to mitigate data sparsity and diversify behavioral patterns. These strategies also help reduce the dominance of frequent behaviors and improve learning from long-tail ones: (1) Co-occurrence behavior addition adds potential co-occurring behaviors based on global statistics to simulate realistic behavior combinations; (2) Frequency-based behavior masking masks frequent behaviors, which encourages the model to focus on more informative rare behaviors; (3) Auxiliary behavior flipping randomly perturbs high-frequency auxiliary behaviors (e.g., clicks), reducing dependence on dominant behaviors and enhancing generalization to diverse behavior combinations. Distinct from item-based augmentations, these methods diversify training views without compromising the semantics of the item sequence. Furthermore, we introduce a behavior richness-based loss weighting, assigning higher loss weights to prediction steps with richer behavior sets to enhance the model's ability to learn from complex supervisory signals.

We summarize our main contributions as follows.
\begin{itemize}
\item We design a dual item-behavior fusion architecture that collaboratively leverages early and intermediate fusion strategies to improve semantic modeling capabilities for multiple behaviors.
\item We propose three behavior-level augmentation methods for BSSR that operate on behavior sequences to generate diverse yet semantically consistent views, alleviating both data sparsity and behavior imbalance.
\item Extensive experiments on three real-world datasets are conducted to verify the effectiveness of BLADE.
\end{itemize}

\section{Related Work}
\paragraph{Single-Behavior Sequential Recommendation.}
Early work on SR employed Markov chains \cite{fpmc,he2016fusing} to model item transitions. With the advancement of deep neural networks, numerous architectures have been explored, including RNN-based \cite{grurec}, CNN-based \cite{caser}, and attention-based models \cite{sasrec,bert4rec,wang2025intent}, which laid the foundation for a range of follow-up improvements. Despite their success, these methods typically model only one type of behavior (e.g., purchases), overlooking auxiliary behaviors such as clicks and favorites. Therefore, they may not adapt well to real-world recommendation scenarios.

\paragraph{Multi-Behavior Sequential Recommendation.}
Most existing studies on MBSR are deep learning-based algorithms, including RNN-based models \cite{dumys,rnn1,rnn2}, GNN-based models \cite{chen2022global,mgnn}, Transformer-based models \cite{zhan2022transrec++,mbstr,nextip} and hybrid techniques-based models \cite{xia2022multi,knowledge,meng2020incorporating}. DyMuS \cite{dumys}, a recent RNN-based framework, models each type of user behavior with an individual GRU, capturing behavior-specific dynamics. It then employs a dynamic routing mechanism to adaptively combine the resulting behavior-aware representations. MBHT \cite{mbht} combines multi-scale Transformers and multi-behavior hypergraph learning to model short-term dynamics at different temporal granularities and long-range dependencies across behavior types, effectively enhancing user preference modeling. MB-STR \cite{mbstr} configures the weights in the classic multi-head self-attention layer to be behavior-specific. It also introduces a relative positional encoding scheme and uses MMoE to integrate behavior representations. To better represent social media behavior patterns, EIDP extends MBSR to the BSSR setting by modeling each interaction as a behavior set \cite{chen2024explicit}, allowing one item to be linked to multiple behaviors. While this richer representation increases expressiveness, BSSR still faces long-standing challenges such as data sparsity and behavior heterogeneity. 

\section{Method}

\begin{figure*}[t] 
    \centering
    \includegraphics[width=\textwidth]{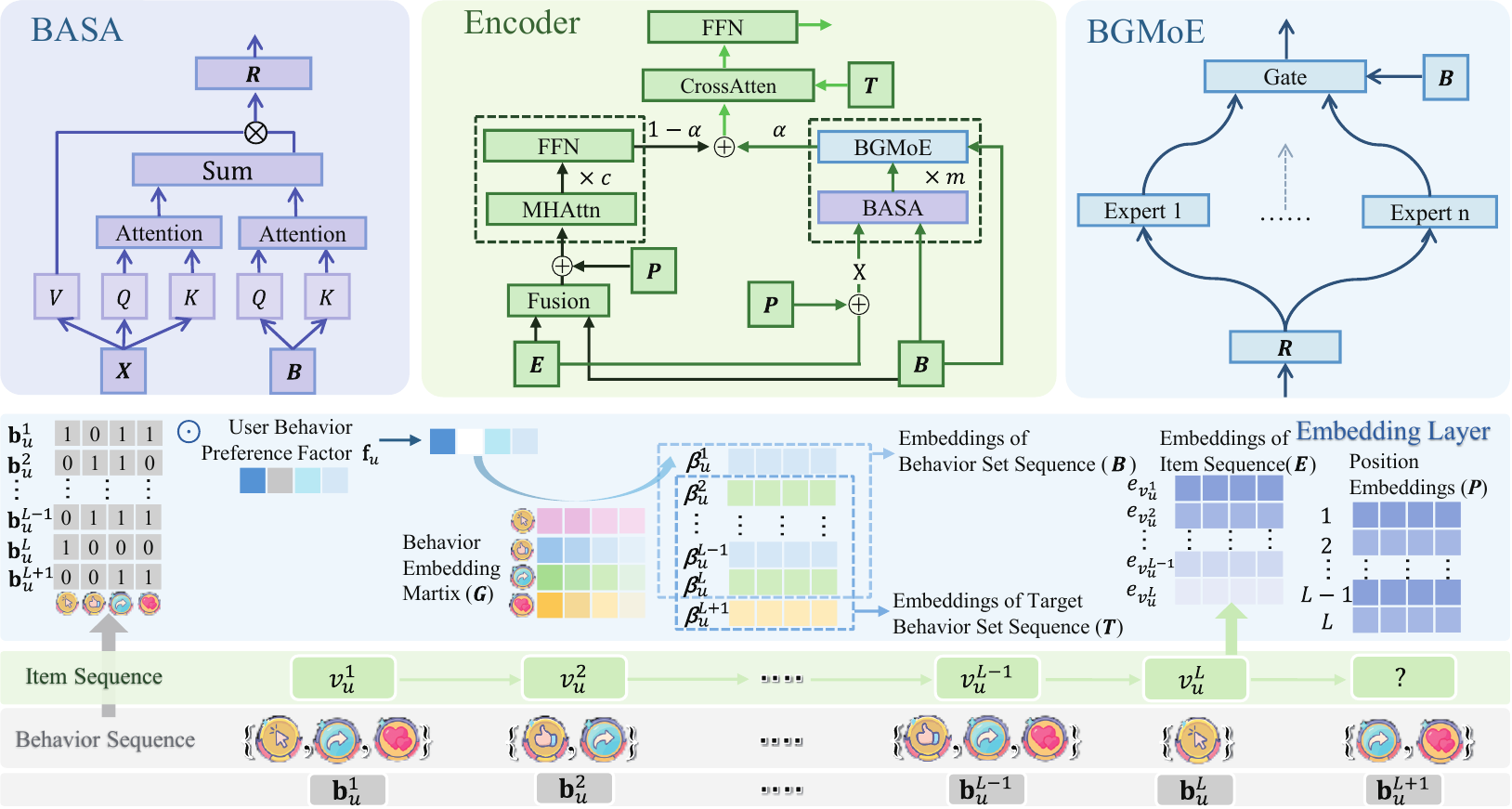}
    \caption{Overview of the proposed dual item-behavior fusion architecture, which integrates behavior information at both the input and intermediate layers. For clarity, residual connections and normalization layers are omitted in the illustration.} \label{fig:main}
\end{figure*}

\subsection{Problem Formulation}
We use $\mathcal{U} = \{u\}$, $\mathcal{V} = \{v\}$ and $\mathcal{B}$ to denote a set of users, a set of items and a set of user behaviors, respectively. $|\mathcal{U}|$, $|\mathcal{V}|$ and $|\mathcal{B}|$ are the numbers of users, items and behavior types, respectively. The interaction sequence with behavior sets of a specific user $u \in \mathcal{U}$ can be represented as: $\mathcal{S}_u = \{(v_u^1, \textbf{b}_u^1), \ldots, (v_u^l, \textbf{b}_u^l), \ldots, (v_u^L, \textbf{b}_u^L)\}.$ We represent the general form of a behavior set as a multi-hot vector, $\textbf{b}_u^l = (b_{u,1}^l, \ldots, b_{u,k}^l, \ldots, b_{u,|\mathcal{B}|}^l) \in \mathbb{R}^{|\mathcal{B}|},$ where $b_{u,k}^l = 1$ if user $u$ has interacted with item $v_u^l$ with the $k$-th behavior at the $l$-th step, and $b_{u,k}^l = 0$ otherwise. $L$ indicates the length of the user sequence. 

The goal is to predict the next item $v \in \mathcal{V} \setminus \mathcal{S}_u$ that is likely to be interacted with under a target behavior set by user $u$ at the $L+1$ step. The target behavior set is a vector containing the target behaviors.
\subsection{Overview of BLADE}
To address behavior heterogeneity and data sparsity in BSSR, we propose BLADE, which comprises two components: (1) dual item-behavior fusion: We integrate item and behavior information via early and intermediate fusion to enhance semantic modeling of behaviors and capture multi-granularity behavioral patterns; (2) behavior-level data augmentation: We design three augmentation methods—Co-occurrence addition, Frequency-based masking, and Auxiliary behavior flipping—that operate at the behavior level while preserving item-sequence semantics.

\subsection{Dual Item-Behavior Fusion}
In our dual item-behavior fusion architecture, we integrate item and behavior information at both early and intermediate stages, enhancing modeling capacity across multiple semantic levels. Figure~\ref{fig:main} illustrates the overall architecture.

\subsubsection{Embedding Layer.} 
An embedding matrix $\bm{V} \in \mathbb{R}^{|\mathcal{V}| \times d}$ maps items to $d$-dimensional vectors. Given a sequence $\mathcal{S}_u$, we obtain its item embeddings via lookup and stack them as $\bm{E} = [\, \bm{e}_{v^1_u} ; \bm{e}_{v^2_u} ;\, \ldots ;\, \bm{e}_{v^L_u} ] \in \mathbb{R}^{L \times d}.$ To encode temporal order, a learnable positional encoding $\bm{P} \in \mathbb{R}^{L \times d}$ is constructed. For behavior set encoding, following the design in EIDP \cite{chen2024explicit}, we distinguish different behavior types by introducing a behavior embedding matrix $\bm{G} \in \mathbb{R}^{|\mathcal{B}| \times d}$. Considering that different users may express varying preference intensities towards different behaviors, a factor matrix $\bm{F} \in \mathbb{R}^{|\mathcal{U}| \times |\mathcal{B}|}$ is used to model user behavioral preferences. For a given behavior set $\textbf{b}_u^l$, the personalized behavior set embedding can be computed as:
{\small
\begin{equation}
    \boldsymbol{\beta}_u^l = \text{softmax} (\textbf{f}_u \odot \textbf{b}_u^l) \cdot \bm{G},
    \label{eq:behavior_encoding}
\end{equation}
}where $\odot$ is the element-wise product, $\textbf{f}_u \in \mathbb{R}^{|\mathcal{B}|}$ denotes the behavioral preference factor of user $u$, and $\boldsymbol{\beta}_u^l \in \mathbb{R}^{d}$ can be viewed as the embedding corresponding to the behavior set $\textbf{b}_u^l$. Note that, we  will use the uppercase symbols $\bm{B}$ and $\mathfrak{B}$ to represent the matrix forms of $\boldsymbol{\beta}$ and \textbf{b}, respectively. For brevity, we omit the user index $u$ in the following derivations when there is no ambiguity.

\subsubsection{Early Item-Behavior Fusion.}
We integrate the item embeddings $\bm{E}$ and the behavior set embeddings $\bm{B}$ at the input layer to enable rich interactions between items and behaviors, as follows:
{\small
\begin{equation}
    \bm{E}' = f(\bm{E}, \bm{B}),
\end{equation}
}where the function $f(\cdot)$ can be instantiated as summation, concatenation, or gating. Then we adopt a widely used self-attentive architecture, i.e., Transformers \cite{transformer}. Specifically, it consists of stacks of multi-head self-attention layers (denoted by MHAttn($\cdot$)) and point-wise feed-forward networks (denoted by FFN($\cdot$)). The input and the output can be formalized as follows:
{\small
\begin{align}
&\bm{X}_e= \bm{E}' + \bm{P}, \\
\bm{F}&= \text{FFN}(\text{MHAttn}(\bm{X}_e)).
\end{align}
}The output $\bm{F} \in \mathbb{R}^{L \times d}$ serves as the contextualized representation after early item-behavior fusion.

\begin{figure*}[t] \centering
    \includegraphics[width=\textwidth]{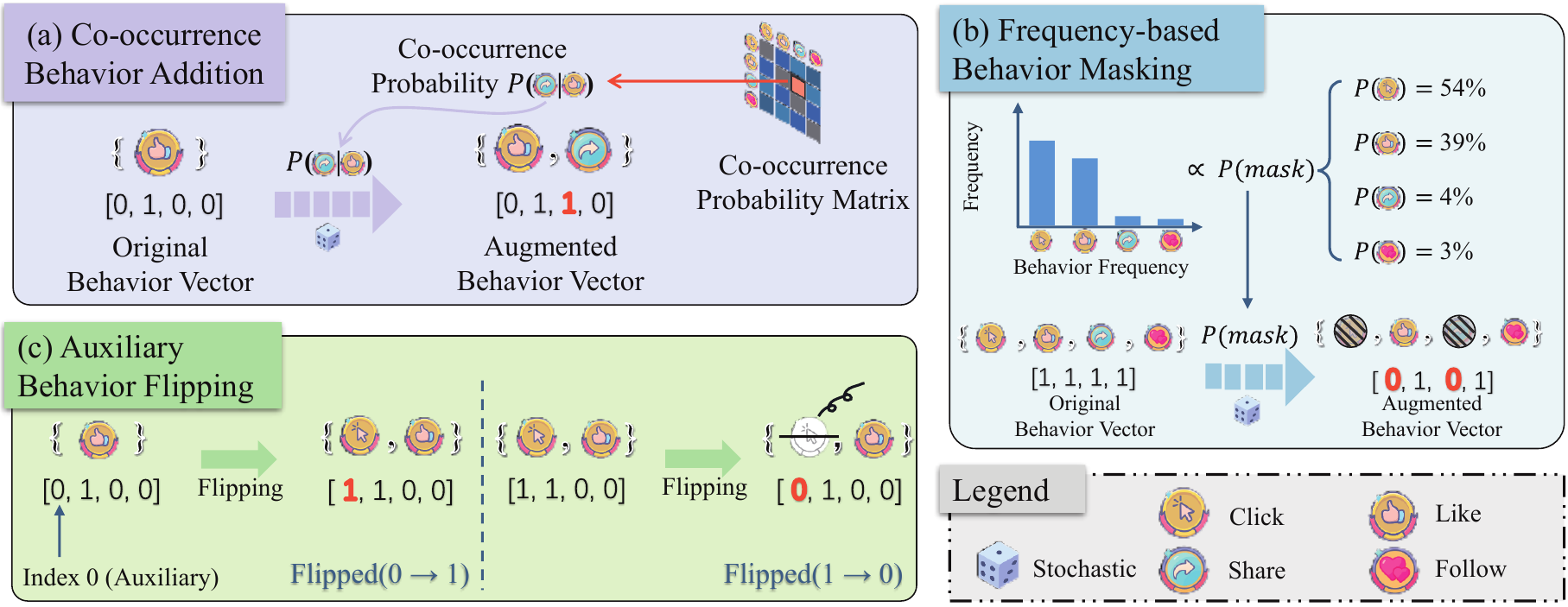}
    \caption{Overview of the proposed behavior-level data augmentation methods: (a) Co-occurrence behavior addition adds an extra behavior based on co-occurrence frequency;
(b) Frequency-based behavior masking masks behaviors according to their occurrence frequency;
(c) Auxiliary behavior flipping removes or adds an auxiliary behavior.} 
    \label{fig:main_aug}
\end{figure*}

\subsubsection{Intermediate Item-Behavior Fusion.}

To better capture behavior semantics, we extend the Transformer with two components: (1) a behavior-aware self-attention (BASA) module, inspired by \cite{diff}, and (2) a behavior-guided mixture-of-experts (BGMoE) module \cite{moe}. The input is constructed by summing the item and position embeddings, i.e., $\bm{X} = \bm{E} + \bm{P}$. Meanwhile, the behavior set embeddings $\bm{B}$ influence the encoding process by implicitly guiding representation learning. In BASA, behavior-informed queries and keys modulate attention scores, assigning higher weights to items associated with similar behavior sets. In BGMoE, expert weights are dynamically computed based on behavior set embeddings, enabling the model to adaptively highlight the representations of diverse behavioral semantics during user modeling.

\textit{BASA.} For the $h$-th head, the behavior-aware attention matrix is computed by projecting $\bm{X}$ and $\bm{B}$ into query/key spaces, respectively:
{\small
\begin{equation}
\bm{A}_h = \bm{Q}_h^{X} (\bm{K}_h^{X})^{\top} + \bm{Q}_h^{B} (\bm{K}_h^{B})^{\top}.
\end{equation}
}Then, the behavior-aware attention output is:
{\small
\begin{equation}
\bm{R}_h = (\text{softmax}\left( \frac{\bm{A}_h}{\sqrt{d_h}} \right)\odot \Delta) \bm{V}_h^X,
\end{equation}
}where $\bm{V}_h^X$ is the value projection from $\bm{X}$ and $\Delta$ is the lower triangular matrix of causality mask. Outputs from all heads are concatenated and projected to form $\bm{R}$.

\textit{BGMoE.} To facilitate the fusion of behavior information, we apply a MoE module where the behavior set embedding guides expert weights. Given the behavior set embeddings $\bm{B}$ and the attention output $\bm{R}$, the output is computed as:
{\small
\begin{equation}
\bm{O} = \sum_{i=1}^n\phi(\bm{B})_ie_i(\bm{R}),
\end{equation}
}where $\phi(\cdot)$ computes routing weights via linear projection followed by softmax normalization, and $e_i(\cdot)$ denotes the $i$-th expert implemented by an FFN.

\subsubsection{User Representation.}
The fused representation $\check{\bm{U}}$ is obtained by aggregating the outputs from early and intermediate fusion as follows:
{\small
\begin{equation}
    \check{\bm{U}} = \alpha\bm{O}+(1-\alpha) \bm{F},
\end{equation}
}where $\check{\bm{U}} \in \mathbb{R}^{L \times d}$ and $\alpha$ denotes the representation aggregating hyperparameter. To align user preferences with the next-step behavior set semantics, we apply a cross-attention mechanism between the fused representation $\check{\bm{U}}$ and the next-step behavior set embeddings $\bm{T}=[\boldsymbol{\beta}^2;\,\ldots ;\,\boldsymbol{\beta}^L,\boldsymbol{\beta}^{L+1}]$. Specifically, we treat the $\bm{T}$ as the query, and the fused representation $\check{\bm{U}}$ as both the key and value to compute the final user representation $\bm{U}$.
{\small
\begin{align}
    &\bm{U} = \text{FFN}(\text{CrossAttn}(\bm{T}, \check{\bm{U}})), \\
    \text{CrossAttn}&(\bm{T}, \check{\bm{U}}) = (\text{softmax} \left( \frac{ \check{\bm{Q}} \check{\bm{K}}^{\top} }{ \sqrt{d} } \right)\odot \Delta) \check{\bm{V}},
\end{align}
}where $\check{\bm{Q}} = \bm{T} \bm{W}_{\check{Q}} \in \mathbb{R}^{L \times d}$, $\check{\bm{K}} = \check{\bm{U}} \bm{W}_{\check{K}} \in \mathbb{R}^{L \times d}$, and $\check{\bm{V}} = \check{\bm{U}} \bm{W}_{\check{V}} \in \mathbb{R}^{L \times d}$. Here, $\bm{W}_{\check{Q}}, \bm{W}_{\check{K}}, \bm{W}_{\check{V}}$ are learnable projection matrices. 

\subsection{Behavior-Level Data Augmentation} 
To mitigate data sparsity, we propose three behavior-level data augmentation methods. These operate on user behavior sequences instead of core item sequences to generate diverse interaction views, thereby enhancing model generalization while preserving the semantics of item sequences. Additionally, all three methods are frequency-aware, tending to add low-frequency behaviors and mask high-frequency ones, thus alleviating the effects of behavioral imbalance.

Given a behavior sequence $\mathfrak{B} = [\textbf{b}^1, \textbf{b}^2, \ldots, \textbf{b}^L]$, we first sample a subset of steps for augmentation based on a predefined operation ratio \( \rho \in (0, 1) \). Let $\mathcal{I} = \{{i_1, i_2, \ldots, i_k}\}$ denote the sampled index set, where $k = \lfloor \rho \cdot L \rfloor$. One of the proposed augmentation methods is then applied to the behavior sets at these positions, yielding an augmented behavior sequence $\mathfrak{B}_{*} = [\textbf{b}^1_*, \textbf{b}^2_*, \ldots, \textbf{b}_*^L]$, where $\textbf{b}_*^l = \textbf{b}^l$ if $l \notin \mathcal{I}$, and otherwise $\textbf{b}^l_*$ is the augmented version of $\textbf{b}^l$. Figure~\ref{fig:main_aug} illustrates the augmentation methods.

\subsubsection{Co-occurrence Behavior Addition.}
To enhance the diversity of behavior combinations and simulate joint behavior patterns (e.g., "like + favorite"), this method supplements the original behavior set with frequently co-occurring yet currently missing behaviors. Specifically, given a co-occurrence probability matrix $\mathbf{M} \in \mathbb{R}^{|\mathcal{B}| \times |\mathcal{B}|}$ and a behavior set $\textbf{b}$, we compute an aggregated co-occurrence vector as:

{\small
\begin{equation}
\mathbf{p} = \textbf{b} \cdot \mathbf{M}.
\end{equation}
}To exclude the influence of already-present behaviors, we set $\mathbf{p}_k = 0$ for all $k$ where $\textbf{b}_k = 1$. Then we normalize $\mathbf{p}$ into a probability distribution:
{\small
\begin{equation}
\mathbf{p} \leftarrow \frac{\mathbf{p}}{\sum_{k=1}^{|\mathcal{B}|} \mathbf{p}_k}.
\end{equation}
}Finally, a new behavior $b^+$ is sampled from the distribution $\mathbf{p}$ and added to the set by setting ${\textbf{b}}_{b^+}= 1$.

\subsubsection{Frequency-based Behavior Masking.}
To prevent the model from overfitting to high-frequency behaviors and thereby weakening its ability to model long-tail behaviors, we dynamically mask frequently occurring behaviors, thus encouraging the model to focus more on informative long-tail behaviors. Specifically, we first compute a behavior frequency vector $\mathbf{m} \in \mathbb{R}^{|\mathcal{B}|}$. For a given behavior set $\textbf{b}$, the masking probability for behavior type $i$ is defined as:

{\small
\begin{equation}
   P(\textbf{b}_i=0)=\frac{\mathbf{m}^c_i}{\sum_{k=1}^{|\mathcal{B}|} \mathbf{m}^c_k},
\end{equation}
}where the exponent $c$ is introduced to smooth the influence of extremely high-frequency behaviors. Each behavior type is then independently masked with its corresponding probability; if selected, we set ${\textbf{b}}_{i}= 0$.

\subsubsection{Auxiliary Behavior Flipping.}
Auxiliary behaviors (e.g., clicks) are common forms of implicit feedback, yet they may introduce noise. To prevent the model from overfitting to such behaviors, we flip the auxiliary behavior $b_a$ in a given behavior set $\textbf{b}$ as follows: 
{\small
\begin{equation}
{\textbf{b}}_{b_a} = 1- \textbf{b}_{b_a}.
\end{equation}
}

\subsection{Prediction and Model Training}
\subsubsection{Preferred Item Prediction.}
In the training phase, predict the next item that user $u$ may interact with under the target behavior set at the ($l$+1)-th step:
{\small
\begin{equation}
    \hat{y}_{l+1,v} = \bm{u}_l \bm{e}_v^\top,
\end{equation}
}where $\bm{u}_l$ represents the user representation $\bm{U}$ at the $l$-th step, $\hat{y}_{l+1,v}$ is a scalar that signifies the probability score of interacting with item $v$.
\subsubsection{Training and Optimization.}
We utilize binary cross-entropy (BCE) loss to optimize our model for the next-item prediction task. Additionally, we introduce a behavior richness-based loss weighting to assign higher loss weights to prediction steps where the target behavior set contains multiple behavior types. Let $w_{u,l} = \frac{\| \textbf{b}_{u,l+1} \|_0}{|\mathcal{B}|}$, which increases the penalty for prediction errors under richer behavior supervision. The formula for the loss is:
{\small
\begin{align}
    \mathcal{L}_{\text{next}} = - \frac{1}{|\delta(v)|} \sum_{u \in \mathcal{U}} \sum_{l=1}^{L} 
    &\delta(v^l_u) w_{u,l}  \Big[ 
     \log \sigma(\hat{y}_{l+1,v^l_u}) \nonumber \\
    +\ & \log \big(1 - \sigma(\hat{y}_{l+1,j}) \big) \Big],
\end{align}
}where $\sigma(\cdot)$ denotes the sigmoid function and subscript $j \in \mathcal{V} \setminus \mathcal{S}_u$ denotes a randomly sampled negative item. $\delta(v^l_u)$ is an indicator function: $\delta(v^l_u) = 1$ if $v^l_u$ is a real item and $\delta(v^l_u) = 0$ if it is a padding item. $|\delta(v)|$ represents the total number of valid ground-truth items across all sequences.

To further enhance representation learning, we introduce a sequence-level contrastive loss to enforce consistency between two augmented views of the same user interaction sequence. Given an original sequence $\mathcal{S}_u$, we generate two augmented sequences $\mathcal{S}_u^{aug1}$ and $\mathcal{S}_u^{aug2}$, which are independently encoded into $\bm{H}_u^1, \bm{H}_u^2 \in \mathbb{R}^{L \times d}$. We then concatenate the representations across all steps to obtain $\bm{h}_u^1, \bm{h}_u^2 \in \mathbb{R}^{L \cdot d}$. The contrastive learning loss is defined as:
{\small
\begin{equation}
\mathcal{L}_{\text{SeqCL}} 
= \mathcal{L}_{\text{CL}}(\bm{h}_u^1,\, \bm{h}_u^2)
+ \mathcal{L}_{\text{CL}}(\bm{h}_u^2,\, \bm{h}_u^1),
\end{equation}
}{\small
\begin{equation}
\mathcal{L}_{\text{CL}}(\bm{h}_u^1,\, \bm{h}_u^2)
= - \log 
\frac{
\exp\big( \text{sim}(\bm{h}_u^1,\, \bm{h}_u^2)/ \tau \big)
}{
\sum_{neg} \exp\big( \text{sim}(\bm{h}_u^1,\, \bm{h}_{neg})/\tau \big)
},
\end{equation}
}where $\text{sim}(\cdot)$ denotes the dot product operation, $\bm{h}_{neg}$ denotes augmented representations from other sequences within the current mini-batch and $\tau$ is a temperature parameter.
The final objective combines both losses:
{\small
\begin{equation}
    \mathcal{L} = \mathcal{L}_{\text{next}} + \lambda \mathcal{L}_{\text{SeqCL}},
\end{equation}
}where $\lambda$ is a balancing hyperparameter that controls the contribution of the contrastive loss.

\begin{table*}[t]
\centering
\setlength{\tabcolsep}{0.8mm}
{\small
\def\arraystretch{0.8} 

\begin{tabular}{l|cccc|cccc|cccc}
\toprule
Dataset & \multicolumn{4}{c|}{\textbf{KuaiSAR}} & \multicolumn{4}{c|}{\textbf{QK-Article}} & \multicolumn{4}{c}{\textbf{QK-Video}} \\
\cmidrule(r{2pt}){1-1}
\cmidrule(l{2pt}r{2pt}){2-5}
\cmidrule(l{2pt}r{2pt}){6-9}
\cmidrule(l{2pt}){10-13}
Metrics & NDCG@5 & HR@5 & NDCG@10 & HR@10 & NDCG@5 & HR@5 & NDCG@10 & HR@10 & NDCG@5 & HR@5 & NDCG@10 & HR@10 \\
\midrule
SASRec      & 0.0115 & 0.0180 & 0.0159 & 0.0320
            & 0.0176 & 0.0289 & 0.0255 & 0.0534 
            & 0.0078 & 0.0126 & 0.0104 & 0.0204 \\
            
CL4SRec     & 0.0096 & 0.0160 & 0.0132 & 0.0276
            & 0.0187 & 0.0303 & 0.0257 & 0.0522
            & 0.0082 & 0.0126 & 0.0109 & 0.0208\\
            
\midrule
DyMuS       & 0.0020 & 0.0039 & 0.0030 & 0.0073
            & 0.0057 & 0.0099 & 0.0075 & 0.0155 
            & 0.0050 & 0.0077 & 0.0072 & 0.0146\\
DyMuS$^{+}$ & 0.0024 & 0.0045 & 0.0036 & 0.0081
            & 0.0082 & 0.0141 & 0.0113 & 0.0238 
            & 0.0033 & 0.0056 & 0.0058 & 0.0133\\
            
MBHT        &0.0110	&0.0181	&0.0151	&0.0312
            &0.0193	&0.0314 &0.0250 &0.0491
            &0.0051&	0.0081&	0.0078&	0.0165\\
            
MB-STR      & 0.0077 & 0.0131 & 0.0109 & 0.0228
            & 0.0156 & 0.0245 & 0.0221 & 0.0449 
            & 0.0080 & 0.0130 & 0.0116 & 0.0243\\
\midrule
EIDP     
 & \underline{0.0122} & \underline{0.0184} & \underline{0.0169} & \underline{0.0331}
 & \underline{0.0198} & \underline{0.0317} & \underline{0.0288} & \underline{0.0599} 
 & \underline{0.0095} & \underline{0.0159} & \textbf{0.0137} & \textbf{0.0291}\\
\midrule
BLADE
& \textbf{0.0135} & \textbf{0.0218} & \textbf{0.0187} & \textbf{0.0380}
& \textbf{0.0215} & \textbf{0.0354} & \textbf{0.0301} & \textbf{0.0621} 
& \textbf{0.0097} & \textbf{0.0161} & \underline{0.0127} & \underline{0.0256} \\
\bottomrule
\end{tabular}
}
\caption{Performance comparison on KuaiSAR, QK-Article and QK-Video datasets with NDCG@5/10 and HR@5/10. The best results are marked in bold, and the second best results are underlined.}
\label{tab:main_results}
\end{table*}

\begin{table}[htbp]
  \centering
\setlength{\tabcolsep}{1mm}
{\small
  \begin{tabular}{lccc}
    \toprule
    \textbf{Dataset} & \textbf{\#Users} & \textbf{\#Items} & \textbf{\#Interactions} \\
    \midrule
    KuaiSAR    & 3,812   & 10,653   & 528,242\\
    \addlinespace
    QK-Article & 5,081   & 13,788   & 252,069\\
    \addlinespace
    QK-Video   & 5,081   & 20,494   & 150,396\\
    \bottomrule
  \end{tabular}
  }
  \caption{Statistics of datasets used in experiments}
  \label{tab:dataset}
\end{table}

\section{Experiments}

\subsection{Experimental Settings}

\subsubsection{Datasets.}
We conduct experiments on the recommendation subset of KuaiSAR \cite{kuaisar} and the Tenrec dataset \cite{tenrec}. i) \textbf{KuaiSAR.} The items are short videos. We treat click as the auxiliary behavior, and like, share and follow as target behaviors. ii) \textbf{QK-Article.} The items are articles. We treat read as the auxiliary behavior, and like, share, favorite and follow as target behaviors. iii) \textbf{QK-Video.} The items are short videos. We treat click as the auxiliary behavior, and like, share and follow as target behaviors. Dataset statistics are summarized in Table \ref{tab:dataset} and the details of data preprocessing are provided in code link.

\subsubsection{Evaluation Metrics.}  We use two widely adopted ranking-oriented evaluation metrics, i.e., hit ratio (HR@$k$) and normalized discounted cumulative gain (NDCG@$k$), where $k \in \{5, 10\}$. We adopt the full-ranking setting in evaluation. 

\subsubsection{Baselines.} We compare BLADE with three categories of baselines: (1) \textbf{Single-Behavior Sequential Recommendation}: SASRec \cite{sasrec} and CL4SRec \cite{cl4rec}; (2) \textbf{Multi-Behavior Sequential Recommendation}: DyMuS \cite{dumys}, MBHT \cite{mbht} and MB-STR \cite{mbstr}; (3) \textbf{Behavior Set-informed Sequential Recommendation}: EIDP \cite{chen2024explicit}. Baseline descriptions are in code link. Following EIDP, we retain only the most preferred behavior in each set based on global frequency, reducing BSSR to a standard MBSR setup. Necessary adaptations for MBSR models are detailed in code link.

\subsubsection{Implementation Details.}
For a fair comparison, we fix the embedding dimension $d$ to $32$ and the sequence length $L$ to $50$ for all models. All models are trained on the same truncated sequences. For BLADE, we tune the number of stacked blocks in both early and intermediate fusion from $\{2, 3\}$, the number of attention heads from $\{2, 4, 8\}$, the dropout rate from $\{0.2, 0.3, 0.4, 0.5\}$, and the number of experts in BGMoE from $\{4, 6, 8\}$. More details are provided in code link.

\subsection{Overall Performance} We compare the proposed model with baseline models across three datasets. Table~\ref{tab:main_results} summarizes the overall performance. Due to space limitations, the results of BLADE reported in Table~1 are the best among the three augmentations. From the results, we summarize the observations as follows: (1) Although MBSR methods introduce additional behavioral information, they do not consistently outperform traditional SR models. In BSSR settings, behavior sets often exhibit higher-order heterogeneous dependencies, which MBSR models—primarily relying on single behavior labels—struggle to capture effectively. Among them, DyMuS performs poorly on all three datasets, likely due to its late fusion strategy that weakly models item-item and behavior-behavior interactions, especially in BSSR settings with richer behavior information. (2) Model performance varies notably across datasets. MBHT excels on KuaiSAR and QK-Article but underperforms on QK-Video, whereas MB-STR shows the opposite trend. This suggests that different models adapt differently to behavioral patterns and scenario-specific characteristics. (3) EIDP, tailored for BSSR, outperforms all MBSR baselines by fully exploiting information within behavior sets, while MBSR methods simplify the behavior set into a single behavior and lose key semantic dependencies. (4) Our BLADE achieves the best performance on most metrics, confirming the effectiveness of dual item-behavior fusion modeling and behavior-level data augmentation in modeling complex user behavioral preferences, alleviating data sparsity, and mitigating behavior distribution imbalance.

\subsection{Ablation Study and Component Analysis}
We validate the effectiveness of the key components of BLADE on three datasets, KuaiSAR, QK-Article, and QK-Video, through the ablation study as shown in Table~\ref{tab:ablation_blade}. (i) The proposed dual item-behavior fusion architecture outperforms variants using only Early Fusion (EF) or Intermediate Fusion (IF), suggesting that the two strategies capture distinct yet complementary aspects of user behavior and their combination enables more comprehensive preference modeling. (ii) Removing the contrastive loss leads to a substantial drop in performance, demonstrating the importance of contrastive learning over augmented user sequences. This result also highlights the effectiveness of our behavior-level data augmentation methods. (iii) The behavior richness-based loss weighting improves performance by assigning higher weights to target items with multiple behaviors. Replacing it with standard BCE results in a performance drop, highlighting the advantage of leveraging richer supervision for next-item prediction.

\begin{table}[htbp]
  \centering
\setlength{\tabcolsep}{0.4mm}
{\small
\def\arraystretch{0.8} 
  \begin{tabular}{c|c|cccc|c}
    \toprule
    Dataset & Metric & w/o EF & w/o IF & w/o CL & w/o BRW & BLADE \\
    \midrule
    \multirow{2}{*}{KuaiSAR} & NDCG@5 & 0.0105& 0.0120& 0.0107& 0.0071 & \textbf{0.0135} \\
    & HR@5 & 0.0157&  0.0199& 0.0171 &0.0128& \textbf{0.0218} \\
    \midrule
    \multirow{2}{*}{QK-Article} & NDCG@5 &0.0215 &0.0215& 0.0189 & 0.0208 & \textbf{0.0215} \\
    & HR@5 &  0.0352& 0.0348 & 0.0318 & 0.0350 & \textbf{0.0354} \\
    \midrule
    \multirow{2}{*}{QK-Video} & NDCG@5 & 0.0091&0.0081 &0.0078& 0.0074 & \textbf{0.0097} \\
    & HR@5 &0.0146& 0.0138 &0.0128 & 0.0120 & \textbf{0.0161} \\
    \bottomrule
  \end{tabular}
  }
  \caption{Ablation study on KuaiSAR, QK-Article and QK-Video datasets, where “w/o” denotes the removal of the corresponding module in BLADE}
  \label{tab:ablation_blade}
\end{table}

\begin{figure}[t]
    \centering
    \includegraphics[width=\linewidth]{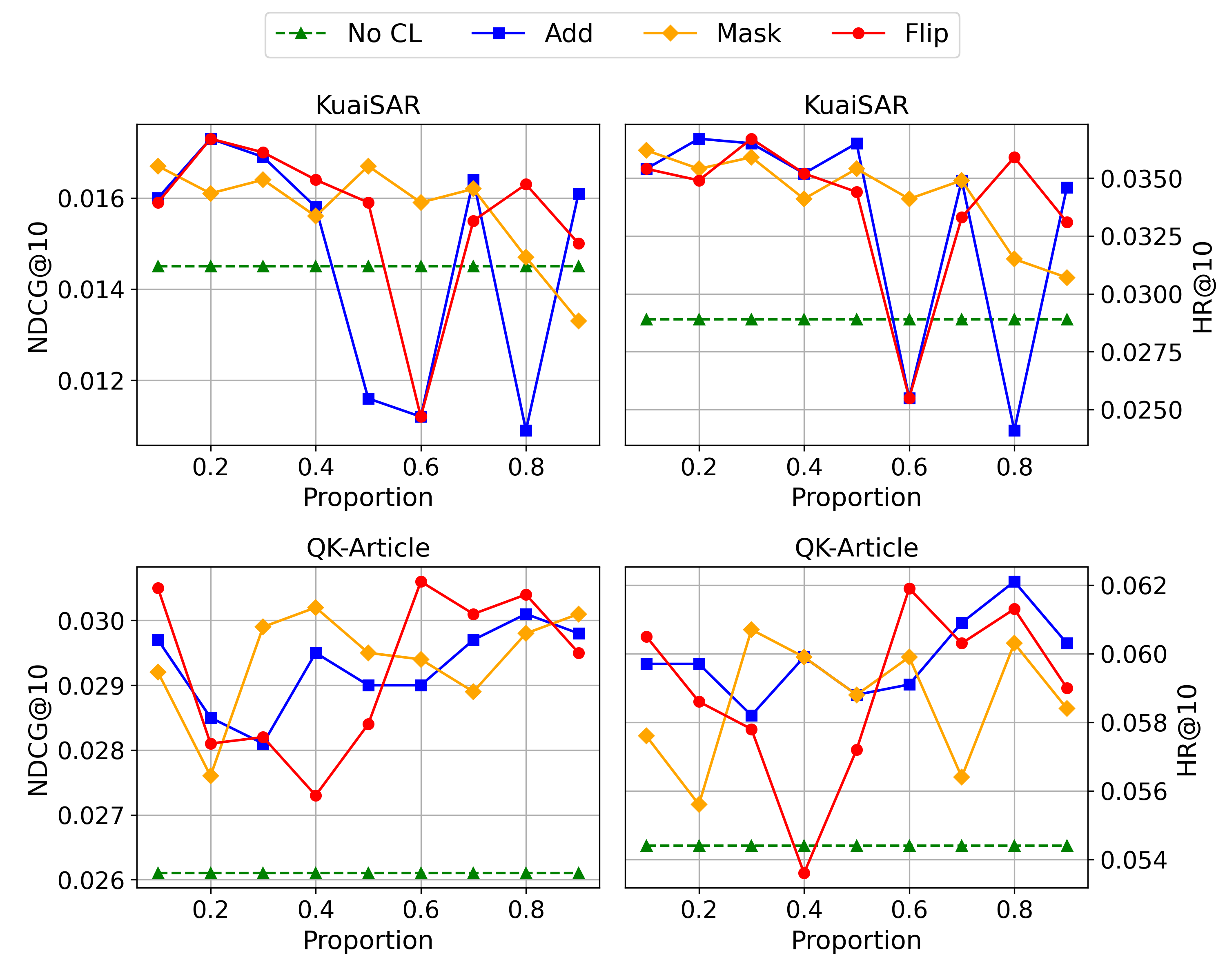}
    \caption{Performance impact of different augmentation methods and operation ratio $p$.}
    \label{fig:aug}
\end{figure}

\subsection{Augmentation Method Comparison} 
\begin{figure}[ht]
    \centering
    \includegraphics[width=\linewidth]{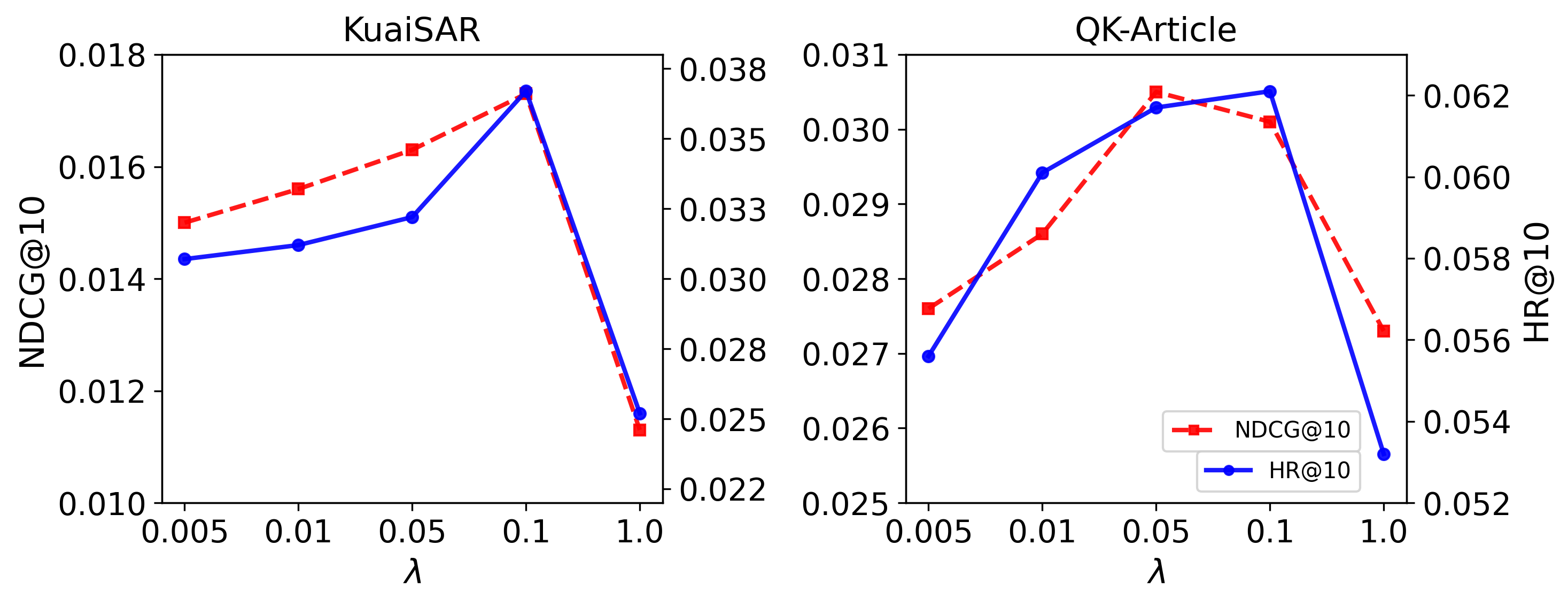}
    \caption{Performance with varying contrastive loss balancing hyperparameter $\lambda$. }
    \label{fig:lamba}
\end{figure}

We analyze how different data augmentation operators and their proportion parameters affect model performance as shown in Figure~\ref{fig:lamba}. To study the effect of each augmentation strategy, we apply only one type of augmentation at a time during contrastive learning. The figure illustrates the performance trends of the three augmentation methods as the proportion parameter p varies from 0.1 to 0.9. We make the following observations: BLADE with any of the proposed augmentation methods consistently outperforms the version without augmentation across most proportion settings and on both datasets. This demonstrates the effectiveness and robustness of our behavior-level data augmentation methods, as they introduce implicit self-supervised signals embedded in the original data and enrich interaction views by operating user behavior sequences rather than core item sequences, thereby significantly enhancing model generalization while preserving item sequence semantics.

\begin{figure}[t]
    \centering
    \includegraphics[width=\linewidth]{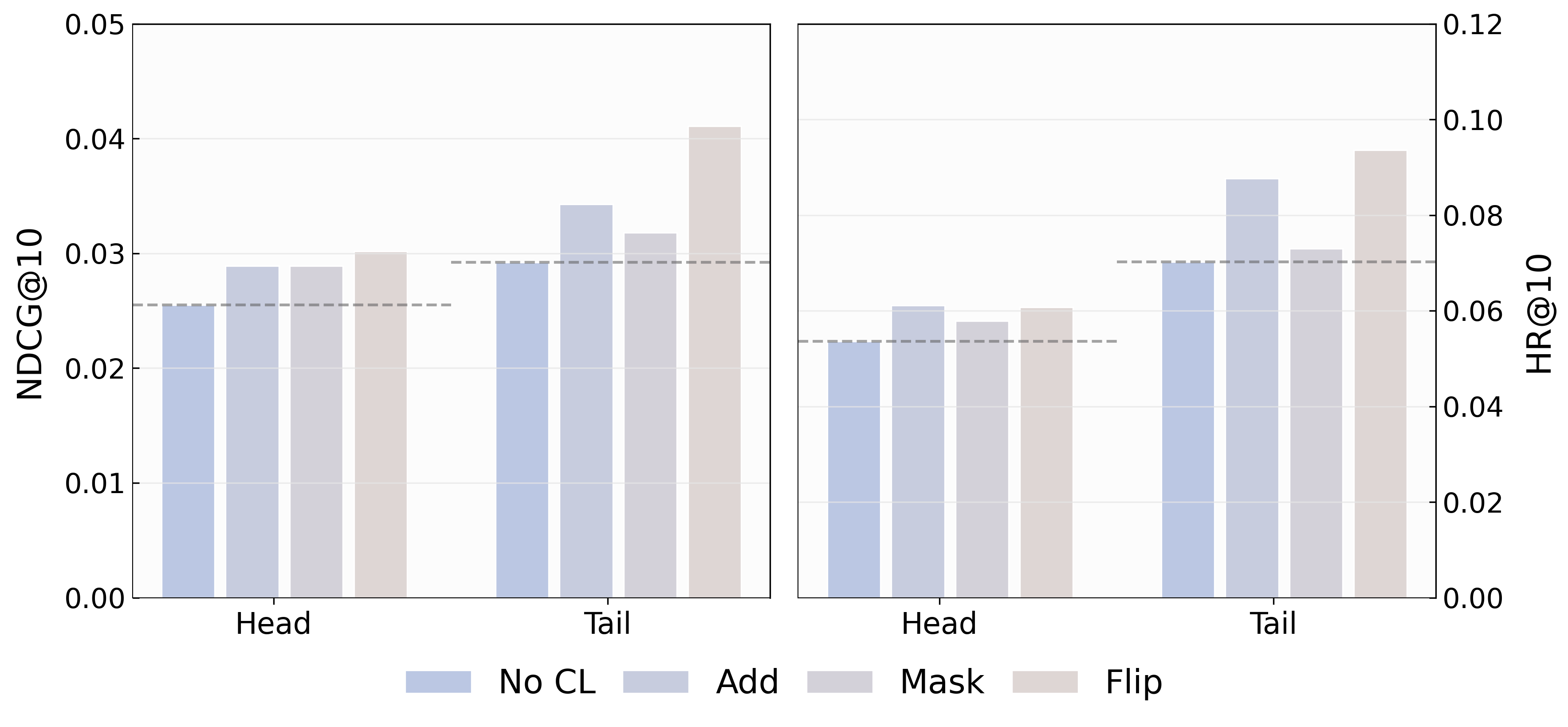}
    \caption{
Performance comparison of three data augmentation methods based on the long-tail degree of user interaction behaviors on the QK-Article dataset.}
    \label{fig:tail}
\end{figure}

\subsection{Impact of Behavior-Level Data Augmentation on Long-Tail Behavior}
In Figure~\ref{fig:tail}, we evaluate the model's performance on the QK-Article dataset by dividing the test user sequences into two groups: the tail group consists of samples where the proportion of long-tail behaviors (share and follow) in users' interaction behaviors exceeds 80\%, while the remaining samples are categorized into the head group. This partition allows us to evaluate the model’s capability in modeling long-tail behaviors. The experimental results show that BLADE consistently outperforms variants without data augmentation and contrastive learning in both groups. Notably, the performance gain is more pronounced in the tail group, indicating that our method is particularly effective at enhancing the model's capacity to model long-tail behaviors. This improvement is mainly due to our frequency-based augmentation methods, which add low-frequency behaviors or mask high-frequency ones, thereby alleviating the imbalance in the original behavior distribution and improving the modeling of long-tail behaviors.

\begin{figure}[ht]
    \centering
    \includegraphics[width=\linewidth]{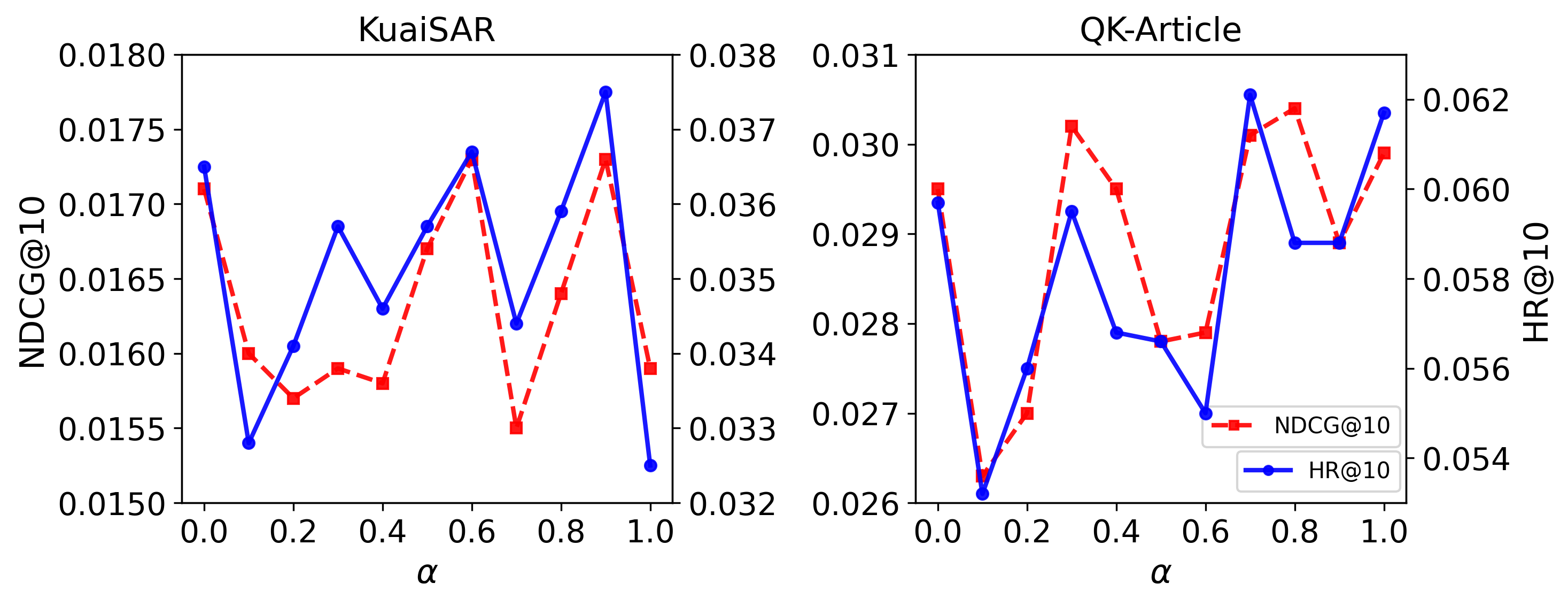}
    \caption{Performance with varying representation aggregation hyperparameter $\alpha$. }
    \label{fig:alpha}
\end{figure}

\subsection{Hyperparameter Sensitivity Analysis}
\textbf{Loss Balancing Hyperparameter.} 
Figure~\ref{fig:lamba} illustrates the impact of the contrastive loss balancing hyperparameter $\lambda$ on the KuaiSAR and QK-Article datasets. The results show that the model achieves the best performance when $\lambda \approx 0.1$, while overly small or large values result in suboptimal results. A small $\lambda$ introduces insufficient contrastive supervision, whereas a large $\lambda$ overemphasizes the contrastive objective, thereby impairing the model’s ability to perform accurate next-item prediction.

\noindent \textbf{Representation Aggregating Hyperparameter.}
Figure~\ref{fig:alpha} reports the effect of the representation aggregating hyperparameter $\alpha$ on model performance across both KuaiSAR and QK-Article. We observe notable performance fluctuations as $\alpha$ varies, clearly suggesting that the model is highly sensitive to $\alpha$. These results indicate that early and intermediate fusion are not strictly complementary. An inappropriate setting of $\alpha$ may lead to suboptimal aggregation of early and intermediate fusion representations.

\section{Conclusion}
In this paper, we propose BLADE, a behavior-level data augmentation framework with dual fusion modeling. To address multi-behavior heterogeneity, BLADE introduces a dual item-behavior fusion architecture that incorporates behavioral information at both the input layer and intermediate layer thereby enhancing semantic representation learning across diverse behaviors. To mitigate data sparsity, BLADE introduces three behavior-level data augmentation methods, which operate on behavior sequences while preserving the semantics of item sequences. Experiments on real-world datasets validate its effectiveness.

\section*{Acknowledgements}
This research was supported by grants from the National Natural Science Foundation of China (No. 62502486), the grants of Provincial Natural Science Foundation of Anhui Province (No. 2408085QF193), USTC ResearchFunds of the DoubleFirst-Class Initiative (No. YD2150002501), the Fundamental Research Funds for the Central Universities of China (No. WK2150110032).

\bibliography{aaai2026}

\end{document}